\documentclass[prl,english,twocolumn,amsmath,showpacs]{revtex4}
\usepackage[latin1]{inputenc}
\usepackage{graphicx}
\usepackage{color}
\usepackage{babel}

\begin{document}

\title{Point-contact spectroscopy of the antiferromagnetic
superconductor HoNi$_{2}$B$_{2}$C in the normal and superconducting
state}

\author{Yu.G.\ Naidyuk, O.E.\ Kvitnitskaya, I.K.\ Yanson}

\affiliation{B.\ Verkin Institute for Low Temperature Physics and Engineering,
National Academy of Sciences of Ukraine, 47 Lenin Ave., 61103, Kharkiv,
Ukraine}

\author{G.\ Fuchs, K.\ Nenkov, A.\ W\"{a}lte, G.\ Behr,
D.\ Souptel, and S.-L. Drechsler }

\affiliation{Leibniz-Institut f\"{u}r Festk\"{o}rper- und
Werkstoffforschung Dresden e.V., Postfach 270116, D-01069 Dresden,
Germany}

\date{\today{}}

\begin{abstract}
Point-contact (PC) spectroscopy measurements on antiferromagnetic
(AF) ($T_{\mbox{\tiny N}}\simeq$5.2\,K) HoNi$_{2}$B$_{2}$C single crystals in
the normal and two different superconducting (SC) states
($T_{c}\simeq$8.5\,K and $T_{c}^{*}\simeq5.6$\,K) are reported.
The PC study of the electron-boson(phonon) interaction (EB(P)I)
spectral function reveals pronounced phonon maxima at 16, 22 and
34\,meV. For the first time the high energy maxima at about
50\,meV and 100\,meV are resolved. Additionally, an admixture of a
crystalline-electric-field (CEF) excitations with a maximum near
10\,meV and a `magnetic` peak near 3\,meV are observed. The
contribution of the 10-meV peak in PC EPI constant $\lambda_{\mbox{\tiny PC}}$
is evaluated as 20-30\%, while contribution of the high energy
modes at 50 and 100\,meV amounts about 10\% for each maxima, so
the superconductivity might be affected by CEF excitations. The SC
gap in HoNi$_{2}$B$_{2}$C exhibits a standard single-band BCS-like
dependence, but vanishes at $T_{c}^{*}\simeq5.6$\,K$<T_{c}$,
with $2\Delta/$k$_{\mbox{\tiny B}}T_{c}^{*}\simeq3.9$. The strong coupling
Eliashberg analysis of the low-temperature SC phase with
$T_{c}^{*}\simeq5.6$ K $\sim$ $T_{\mbox{\tiny N}}$, coexisting with the
commensurate AF structure, suggests a sizable value of
the EPI constant $\lambda_{s}\sim$ 0.93. We also provide strong support for the
recently proposed by us  ''Fermi surface (FS) separation''
scenario for the coexistence of magnetism and superconductivity in
magnetic borocarbides, namely, that the superconductivity in the
commensurate AF phase survives at a special (nearly isotropic) FS
sheet without an admixture of Ho 5d states. Above $T_{c}^{*}$ the SC
features in the PC characteristics are strongly suppressed
pointing to a specific weakened SC state between $T_{c}^{*}$ and $T_{c}$.


\end{abstract}
\maketitle

\section{Introduction}

Borocarbide superconductors \textit{RT}$_{2}$B$_{2}$C, where
\textit{R} is a rare-earth element and \textit{T} is a transition
metal element, mainly Ni, (see, e.\,g.,\  Refs.\ \onlinecite{Muller,Budko06}
and further Refs.\ therein) have been studied intensively during the
last decade. However, in spite of their much more simple, less
correlated, 3D electronic structure compared with high-$T_{c}$
materials, the superconductivity in borocarbides is still under debate.
Based on the interpretation of many magnetic, muon spin resonance,
spectroscopic, specific heat, etc. measurements \cite{Muller}
it is widely accepted that the superconducting (SC) state has an
$s$-wave symmetry and the pairing is mediated by the
electron-phonon (el-ph) interaction. On the
other hand for the magnetically ordered borocarbide
superconductors the role of the crystal-electric-field (CEF)
splitting of the 4$f$-shell states of the $R^{+3}$ ion and their
excitations are of fundamental importance to understand both
ordering phenomena. Additionally, the el-ph interaction responsible for
the SC state in these compounds has also been analyzed within the multiband
Eliashberg theory \cite{Shulga0,Drechsler} with emphasis on the
complex Fermi surface with different contribution to the SC state
by different orbitals on different Fermi surface sheets
\cite{Drechsler,drechsler01,drechsler03,drechsler04,Shorikov}.

In the family of borocarbide superconductors HoNi$_{2}$B$_{2}$C is
distinguished: it exhibits a remarkable reentrant
superconductivity and a variety of magnetic orderings
\cite{Muller}. The reentrant behavior occurs in a temperature
range just above the commensurate antiferromagnetic order at about
$T_{\mbox{\tiny \tiny N}}\simeq$5.2\,K and below the
temperature $T_{\mbox{\tiny M}}\simeq6$ K, which separates the
low-temperature magnetic commensurate phase from
the high-temperature paramagnetic phase by a not completely resolved
yet intermediate incommensurate magnetic phase. In the same temperature
range between $T_{\mbox{\tiny N}}$ and $T_{\mbox{\tiny M}}$
incommensurate structures do occur:  an \emph{a}-axis-modulated one
and a spiral structure along the \emph{c}-axis.

By point-contact (PC) spectroscopic investigations \cite{Naid}
both the SC order parameter and the PC electron-boson(phonon)
interaction (EB(P)I) spectral function
$\alpha_{\rm{\mbox{\tiny PC}}}^{2}F(\omega)$ can be determined in studying the
first and second derivatives of the $I(V)$ characteristic of PC`s.
The measurement of the second derivative of the $I(V)$ of PC`s
provides straightforward information on the PC EB(P)I function
$\alpha^{2}F(\omega)$ \cite{Naid}. The knowledge of
$\alpha^{2}F(\omega)$ for conducting systems is a touchstone  to the
phonon-mediated superconductivity which is governed
mainly by the value of the el-ph-coupling parameter
$\lambda=2\int\alpha^{2}F(\omega)\omega^{-1}d\omega$. Moreover,
the comparison of the experimentally determined
$\alpha^{2}_{\rm{\mbox{\tiny PC}}}F(\omega)$ with
the results of model calculations could be used to
distinguish between different scenarios. Thus the PC
spectroscopy could be helpful to illuminate many details of the EPI in
HoNi$_{2}$B$_{2}$C as well as to resolve other possible quasiparticle
interactions, e.\,g., mediated by magnons and CEF excitations.
In addition to this the SC
gap determines the behavior of the $I(V)$ curve for the PC in the
energy range of a few meV. Hence, this region is widely used to determine the
SC gap from a routine fitting by the well-known
Blonder-Tinkham-Klapwijk (BTK) equations \cite{BTK82}. All this
potential information should be used to deepen the present
sparse knowledge of the SC properties and to help to elucidate
the mechanism(s) of superconductivity of the title material.

Up to now, HoNi$_{2}$B$_{2}$C has been investigated by PC
spectroscopy in \cite{Ryba1,Ryba2,Andreone,Yanson1}. The first
three papers are devoted to investigations of the SC gap. In
general, standard isotropic BCS (Eliashberg)-type behavior of the
gap was found below T$_{N}$, while between $T_{\mbox{\tiny N}}$ and
$T_{c}\simeq$8.5\,K the SC features in PC spectra are strongly
suppressed and speculation about some different SC state
\cite{Ryba1} or `gapless` superconductivity \cite{Ryba2} were
undertaken. The PC EBI spectra for HoNi$_{2}$B$_{2}$C were studied
and analyzed in Ref.\cite{Yanson1}. There, the main attention was
devoted to the study of the low energy part of the spectra and to
the investigation and understanding of the so called `soft` mode
at about 4\,meV in the electron-quasiparticle spectrum. However,
the measured spectra were featureless above 20\,meV, although a
number of pronounced phonon peaks are well resolved at  higher
energies by neutron spectroscopy \cite{Gompf}. In this study we
present more detailed PC EBI spectra of HoNi$_{2}$B$_{2}$C in
comparison with those mentioned in the Refs. given above including also
our recent data presented at two international conferences SCES`04 and M2S-HTSC-VIII
\cite{NaidSCES,M2Sconf}. Additionally, to explain our findings with respect
to the SC gap we provide support for the separation of magnetism and
superconductivity on different Fermi surface sheets in magnetic
borocarbides \cite{drechsler01,drechsler03,Shorikov}.

\section{Experimental details}

\begin{figure}
\begin{center}
\includegraphics[width=9.5cm,angle=0]{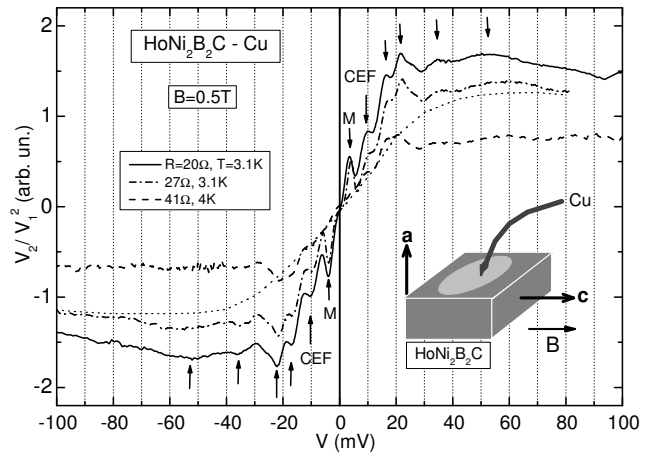}
\end{center}
\caption{Experimental PC spectra {[}see Eq.\,(\ref{pcs1}){]} of
several HoNi$_{2}$B$_{2}$C--Cu PCs. The dotted curve shows an example
of the tentative behavior of the background for the dash-dot curve
spectrum.  A magnetic field of 0.5\,T is applied to suppress
superconductivity. The arrows mark a `magnetic` (M) peak at about
$\pm$3\,mV, a CEF peak, and positions of other resolved peculiarities on the curves
maxima. The inset shows a sketch of the PC configuration (see also the text for more details).}
 \label{hof1}
\end{figure}

We have used single crystals of HoNi$_{2}$B$_{2}$C grown by a
floating zone technique with optical heating \cite{Souptel}. The
residual resistivity is $\rho_{0}\simeq3\mu\Omega$\,cm and the
residual resistivity ratio RRR=13. Final series of measurements
were done for sample with improved quality with RRR=19. The sample
becomes superconducting at about 8.5\,K. PCs were established both
along the c axis and in perpendicular direction by standard
`needle-anvil` or `shear` methods \cite{Naid}. As a counter
electrode Cu or Ag thin wires ($\oslash$=0.15\,mm) were used. The
spectra with resolved phonon features like in Fig.\,\ref{hof1}
were registered by touching of the sharpened Cu wire on a shallow
bright caving of about 1$\times$1mm$^{2}$ on the
HoNi$_{2}$B$_{2}$C crystal surface (see inset in
Fig.\,\ref{hof1}). The electron probe microanalysis (WDX mode) has
shown that the cavity surface has single phase microstructure and
stoichiometric composition well corresponding to the 1:2:2:1
ratio, while excess of Ni has been observed in other places on that
crystal surface.

\section{Results and discussion}

According to the theory of PC spectroscopy \cite{KOS} the second
derivative $R^{-1}dR/dV=R^{-2}d^{2}V/dI^{2}(V)$ of the $I-V$ curve
of the ballistic contact at low temperatures, where $R=dV/dI$, is
determined by the PC EPI function $\alpha_{\rm{\mbox{\tiny PC}}}^{2}F(\omega)$,
\begin{equation}
\label{pcs} R^{\rm -1}\frac{{\rm d}R}{{\rm d}V}= \frac{8\,{\rm
e}d}{3\,\hbar v_{\rm F}}\alpha_{\rm{\mbox{\tiny PC}}}^2(\omega)\,F(\omega)|_{\hbar\omega={\rm e}V} ,
\end{equation}
where $d$ is the PC diameter, e is the electron charge and
$\alpha_{\rm{\mbox{\tiny PC}}}$, roughly speaking, measures the interaction of
an electron with the phonon branches. This interaction is affected
by large angle scattering (back-scattering) processes \cite{KOS}
of electrons in the constriction of the PC. Thus
$\alpha_{\rm{\mbox{\tiny PC}}}^{2}F(\omega)$ is a kind of transport EPI
functions which select phonons with a large momentum or umklapp
scattering. In practice $\alpha_{\rm{\mbox{\tiny PC}}}^{2}(\omega)\, F(\omega)$
can be extracted from the measured rms signal of the second
$V_{2}$ ($V_{2}\propto$ d$^{2}V$/d$I^{2}$) harmonic of a small
alternating voltage $V_{1}$ ($V_{1}\propto$ d$V$/d$I$ is the first
harmonic signal) superimposed on the ramped dc voltage $V$, so
that from (\ref{pcs}):
\begin{equation}
\label{pcs1}
\alpha_{\rm{\mbox{\tiny PC}}}^{2}(\omega)\,F(\omega)|_{\hbar\omega={\rm e}V}=
\frac{3}{2\sqrt{2}}\frac{\hbar v_{\rm F}}{{\rm
e}d}\frac{V_{2}}{V_{1}^{2}}.
\end{equation}
In the case of a heterocontact between two metals the PC spectrum
represents a sum of the contributions from both metals 1 and 2
weighted by the inverse Fermi velocities \cite{Shekhter83}:
\begin{equation}
\frac{V_{2}}{V_{1}^{2}}\propto{R^{-1}}\frac{{\textrm{d}}R}{{\textrm{d}}V}
\propto\upsilon\frac{(\alpha^{2}F)_{1}}{v_{F1}}+(1-\upsilon)
\frac{(\alpha^{2}F)_{2}}{v_{F2}},
\label{het}
\end{equation}
where $\upsilon$ is the relative volume occupied by metal 1 in the PC.
Thus, using of the heterocontacts enables also us to estimate
the relative strength of the EBI in the investigated material as
compared to some standard or well known case.

To study the SC gap we have measured $dV/dI$ characteristics of
N-c-S PCs (here N denotes a normal metal, c is the constriction and S
is the superconductor under study). Then utilizing the generally
used Blonder-Tinkham-Klapwijk equations \cite{BTK82} fits of the
measured curves have been performed. As fit parameters the SC
gap $\Delta $ and the dimensionless barrier parameter $Z$ have been used.
A possible additional smearing of the d$V/$d$I$ curves was also taken
into account by including the parameter $\Gamma$, usually
interpreted as a broadening of the quasiparticle DOS in the
superconductor due to finite life-time effects.

\subsection{PCS of quasiparticle excitations}

\begin{figure}
\begin{center}
\includegraphics[width=8cm,angle=0]{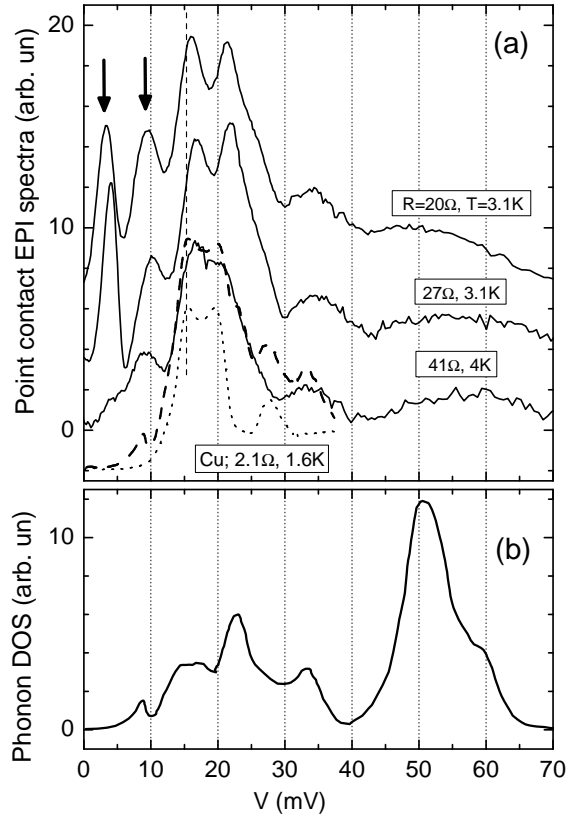}
\end{center}
\caption{(a) PC spectra from the Fig.\,\ref{hof1} with subtracted
background (3 upper solid curves). The arrows mark
the position of the CEF peak and the `magnetic` peak at about 4\,mV,
best visible for the two upper curves. The bottom dotted curve shows the PC spectrum of Cu.
The vertical dashed line marks the position of the first maximum in the Cu spectrum. The dashed
curve presents sum of the PC spectrum of Cu with a function scaled by
the phonon DOS from  panel (b). (b) The neutron phonon DOS for LuNi$_{2}$B$_{2}$C
\cite{Gompf}.} \label{hof2}
\end{figure}

Our PC spectra of HoNi$_{2}$B$_{2}$C demonstrate clear maxima at about
3, 10, 16 and 22\,mV, a smeared maximum at 34\,mV, and a hump
around 50\,mV (Fig.\,\ref{hof1}). The maxima above 15\,mV
correspond well to the phonon DOS maxima of the
nonmagnetic sister compound LuNi$_{2}$B$_{2}$C \cite{Gompf}
[see Fig.\,\ref{hof2}(b)] derived from
inelastic neutron scattering data. Only the high energy part of the
PC spectrum is remarkably smeared. Fig.\,\ref{hof2}(a) displays
the PC EBI function, obtained from the PC spectrum after
subtraction of the background. According to the neutron data
\cite{Gompf} there is a gap around 40\,meV in the phonon DOS,
which separates the acoustic and the optic branches. In this energy region
a minimum occurs in our PC spectra.

\begin{figure}
\includegraphics[width=9.5cm,angle=0]{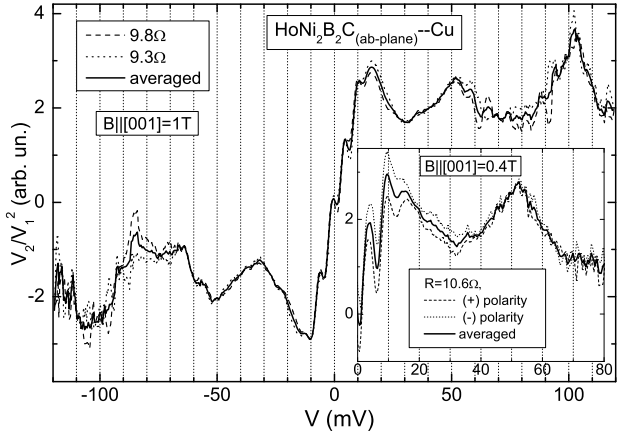}
\caption{PC spectra of HoNi$_{2}$B$_{2}$C--Cu contact with well
resolved high energy maxima around 50 and 100\,mV. PC resistance
was decreased from 10.6$\Omega$ (spectrum in the insert) to
9.3$\Omega$ (main panel) between successive measurements. In the
main panel also averaged for two contacts spectrum (solid line) is
shown, while in the inset averaged for the positive and negative
bias spectrum of the same contact is shown by solid line.
Temperature is 3\,K and magnetic field is applied to suppress
superconductivity.} \label{hof2a}
\end{figure}

We have also succeeded to receive PC spectra with well resolved high
energy maxima around 50 and 100\,mV (see Fig.\,\ref{hof2a}). The
position of the 50-mV maximum coincides with the maximum in the
phonon DOS for LuNi$_{2}$B$_{2}$C shown in Fig.\,\ref{hof2}b, even
the shoulder at about 60\,mV is resolved in the PC spectra (this
feature is more clear at the negative bias). In the phonon DOS
for the nonmagnetic sister compound YNi$_{2}$B$_{2}$C a maximum at around
100\,meV has been observed in Ref.\ \onlinecite{Gompf}. It has been attributed to the
high-frequency B-C bond stretching vibration in accord with
theoretical results obtained in frozen phonon calculations at 106 meV
for the zone-center Raman active mode of YNi$_{2}$B$_{2}$C \cite{Rosner01}.
The low-energy part ($<30$\,meV) of the PC spectra in
Fig.\,\ref{hof2a} shows a less detailed structure as compared to
the spectra from Fig.\,\ref{hof1}. Here maxima only at about 4, 10
and 15\,mV are seen. Also a shoulder in the range of 22\,meV peak is
visible which is more pronounced at negative polarity and
as shown for the spectra in the inset.
However, in interpreting these spectra we have to take into account for
possible contribution of Cu  derived phonons since Cu has been
used as a counter electrode (needle).
The dashed curve in Fig.\,\ref{hof2}(a) presents the sum of the Cu
PC spectrum and the phonon DOS from Fig.\,\ref{hof2}(b). It is
seen that contributions from Cu may be appreciable for the
spectrum of the 41\,$\Omega$ contact in the range between
15--20\,meV, while the two other spectra have pronounced maxima at
16 and 22\,meV, i.~e. at energies higher than the maxima in the Cu
spectrum at 15  and 20\,meV. Moreover, all HoNi$_{2}$B$_{2}$C--Cu
spectra contain no visible contribution from the weaker Cu peak at
about 27\,meV, therefore we conclude that a contribution of Cu is
unsubstantial at least for the two upper spectra in
Fig.\,\ref{hof2}(a). Apparently, the low Fermi velocity in nickel
borocarbides \cite{Muller} as compared to the noble metals
accentuate the intensity of HoNi$_{2}$B$_{2}$C features according
to Eq.\,(\ref{het}). Contrary, the PC spectra in Fig.\,\ref{hof2a}
have likely a contribution from Cu derived phonons in the region
15--20\,mV, which masks the expected HoNi$_{2}$B$_{2}$C phonon maxima so
that the 22\,meV maximum looks like a broad shoulder. In
Fig.\,\ref{hof2a}(inset) also the asymmetry of the spectra versus
bias voltage polarity is clearly seen in the energy region of Cu
phonons, which is characteristic for heterocontact, if the
contribution from both electrodes to the spectrum is comparable \cite{Naid}.

\begin{figure}
\begin{center}
\includegraphics[width=8cm,angle=0]{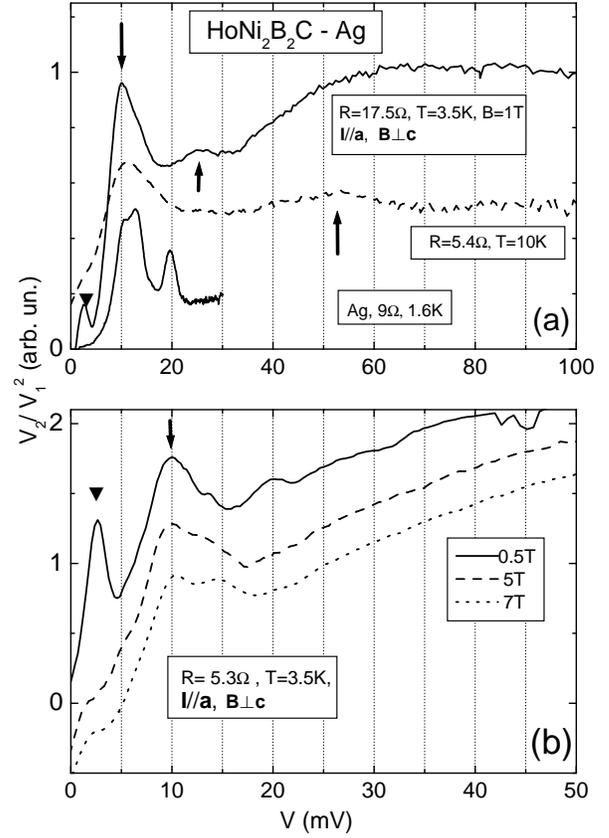}
\end{center}
\caption{PC spectra {[}see Eq.\,(\ref{pcs1}){]} of several
HoNi$_{2}$B$_{2}$C--Ag PCs with a dominant  10-mV peak
(the $\downarrow$ marks the CEF peak, the $\uparrow$ mark the position of the main
phonon peaks from Fig.\ref{hof2}b, the black triangle marks the `magnetic`
peak at about 4\,meV): (a) two spectra with a dominant 10-meV peak. In the first
case the superconductivity is suppressed by magnetic fields,
in the second case -- by high temperature. The latter curve is measured on
a cleaved small piece of a HoNi$_{2}$B$_{2}$C single crystal. The bottom
curve shows the PC spectrum of Ag. (b) The modification of
the spectrum with a dominant 10-mV peak for various applied magnetic fields.} \label{hof3}
\end{figure}

The most prevalent type of HoNi$_{2}$B$_{2}$C PC spectrum  is shown in
Fig.\,\ref{hof3}. Here a maximum close to 3\,mV and
another one around 10\,meV are dominant. The latter maximum might be
connected with CEF excitations, observed in this range by inelastic neutron
scattering studies \cite{Gasser,Cavadini,Kreyssig}. The maximum
around 3\,meV is associated with a magnetically ordered state, since
it disappears above the N\'{e}el temperature mainly in the range
8--10\,K (see dashed curve in Fig.\,\ref{hof3}(a)) and it can be
suppressed by a magnetic field of a few Tesla (see
Fig.\,\ref{hof3}(b)). A detailed discussion of the origin of this
peak will be given in the next section. At the energy of about
10\,meV a contribution from the first peak in the phonon DOS could
also be expected. For LuNi$_{2}$B$_{2}$C the first phonon peak is
observed at about 9\,meV (see Fig.\,\ref{hof2}(b)). For the
lighter (about 6\%) Ho ions the peak position may be somewhat
shifted to higher energies. Note also that the measured
phonon dispersion curves $\omega(q)$ for
HoNi$_{2}$B$_{2}$C \cite{Kreyssig0} have a tendency  that
$\partial\omega(q)/\partial q\rightarrow0 $ around 10--12\,meV
pointing to an enhanced phonon DOS (or maximum) in this
region. However, we have observed that the 10-mV maximum is
considerably affected by a magnetic field (see Fig.\,\ref{hof3}(b)),
therefore a contribution of CEF excitations to the 10-meV peak seems
to be appreciable. Unfortunately, due to `magnetic forces` it was
difficult to keep the mechanical contact stable in  magnetic
fields above 5\,T, especially when the field direction was out of
the $c$-axis. Therefore the influence of the magnetic field on the
10-meV peak (to separate CEF and phonon contribution to this peak)
could be investigated only qualitatively so far.

The spectra discussed here demonstrate weak and smeared phonon
maxima above 10\,mV (see arrows in Fig.\,\ref{hof3}(a)) in
contrast to the spectra shown in Fig.\,\ref{hof2}(a), but they
have a pronounced 10-mV peak (see Fig.\,\ref{hof3}), which points
to the importance of CEF excitations in the transport properties
of HoNi$_{2}$B$_{2}$C. Besides, the electron-CEF interaction may be also
important for a complete understanding of the mechanism of
superconductivity in this compound. We emphasize that
the spectra with the dominant 10-mV peak were qualitatively the
same independent of the counter electrode (Ag or Cu) which
discards remarkable contribution to the PC spectra from these
metals (see the PC spectra of Cu and Ag in Figs.\,\ref{hof2}a and
\ref{hof3}a).

From the measured EBI function (see Fig.\,\ref{hof2}(a)) the EBI parameter
$\lambda_{\rm{\mbox{\tiny PC}}}=2\int\alpha^{2}_{\rm{\mbox{\tiny PC}}}F(\omega)\omega^{-1}d\omega$
was calculated, which was found to be smaller than 0.1 only. However,
there are large uncertainties in the determination of
$\lambda_{\textrm{\tiny PC}}$ if heterocontacts are used. First, this is
because the relative volume $\upsilon$ occupied by the
investigated sample in a PC can deviate from 1/2 [see (\ref{het})].
Secondly, the deviation from the ballistic regime in PCs has to be
corrected by a pre-factor $l_{i}/d$ in (\ref{het}), where $l_{i}$
is the elastic electron mean free path and $d$ is the PC diameter.
However, $l_{i}$ is difficult to evaluate for PC. Further,
the available PC theory and all its equations such as Eq.\,(\ref{pcs1})
have been derived in the framework of the free electron model and
a single-band Fermi surface. Then, paying
attention that the contribution of Cu in the PC spectra of
HoNi$_{2}$B$_{2}$C-Cu heterocontacts in Fig.\,\ref{hof2}(a) is
hardly to be resolved we can conclude that the intensity of the EBI
function in HoNi$_{2}$B$_{2}$C  has to be larger than that in Cu, where
$\lambda_{\textrm{\tiny PC}} \simeq$0.25 \cite{Naid}. This
provides a complementary confirmation of the moderate strength of the EBI in
HoNi$_{2}$B$_{2}$C with a lower bound of 0.25 for
$\lambda_{\textrm{\tiny PC}}$ \cite{lambda}. Moreover the 10-mV peak
contributes about 20-30\% to the $\lambda$ value indicating that
CEF excitations should be taken into account to understand the SC
properties of HoNi$_{2}$B$_{2}$C. The contribution of the high
frequency modes to $\lambda$ can be estimated from the spectra in
Fig.\,\ref{hof2a}. It amounts about 10\% for each of the 50 and 100\,mV
maxima. Note, that by estimating $\lambda$ we discard the
contribution of the `magnetic` peak around 3\,mV, the nature of
which is under discussion (see below).

\begin{figure}
\begin{center}
\includegraphics[width=8cm,angle=0]{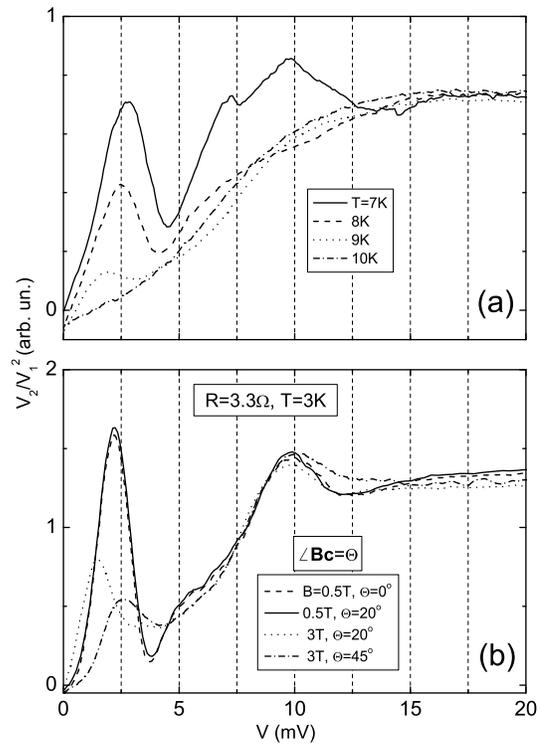}
\end{center}
\caption{(a) PC spectra {[}see Eq.\,(\ref{pcs1}){]} of
HoNi$_{2}$B$_{2}$C--Cu contact established along the a-axis at
different temperature. (b) PC spectra of HoNi$_{2}$B$_{2}$C--Ag
contact established along the a-axis with varying of the magnetic
filed direction relative to the $c$-axis.}
 \label{hof4}
\end{figure}

\subsection{`Magnetic` peak}

Let us turn to the low energy maximum around
e$V_{\mbox{\tiny M}} \approx3$ meV. As we
mentioned above it is very probably related to a
magnetic ordering, since it disappears at temperatures
$T_{\mbox{\tiny M}}\sim$ 8--10\,K (see Fig.\,\ref{hof4}(a))
and it can be suppressed by an external magnetic field (Fig.\,\ref{hof4}(b)).
Figure\,\ref{hof4}(b) shows that by increasing the angle between
the applied magnetic field and the $c$-axis, the height  of the
maximum goes down and its position changes. Thus, this maximum is
more robust, if the field is applied along the $c$-axis.
This is in line with the result reported in Ref.\ \onlinecite{Rathna} that for
a field along the $c$-axis the antiferromagnetic phase with
$T_{\textrm{\tiny N}}\simeq5$\,K is robust up to quite large fields, while
for the perpendicular directions this antiferromagnetic phase is
already suppressed above $B$=0.5\,T at $T$=3\,K. For our PCs both
$T_{\textrm{\mbox{\tiny M}}}$ and the magnetic field which suppresses
the `magnetic` peak are higher. In particular, especially the magnetic
field is enhanced which reaches here up to 5\,T (see
Fig.\,\ref{hof3}(b)). In this context it  should be mentioned that
inside a PC a remarkable (uniaxial) pressure can be created. This might  cause
an  increase of $T_{\textrm{\tiny N}}$ with a ratio up to
d$T_{\textrm{\tiny N}}$/d$P\simeq$8\,K/GPa \cite{Koba}.

For completeness it should be mentioned that spectra with a dominant
3 meV peak (its position may vary between 2.5 and 5 meV) and the
practically absence of any other maxima were observed on
uncleaved `as-grown` surfaces of HoNi$_{2}$B$_{2}$C.
Later on the electron probe microanalysis showed deviations from the
stoichiometry on this surface, which can modify both
T$_{\mbox{\tiny N}}$ and the characteristic magnetic field. As mentioned in
Ref.\ \onlinecite{Souptel} crystallographically oriented thin plates of the
nonsuperconducting HoNiBC compound have been observed in selected parts of
as-grown HoNi$_{2}$B$_{2}$C crystals. It is well-known that the 1:1:1:1 compound
HoNiBC exhibits long-range magnetic order below $T_{\textrm{\tiny N}}\simeq$10\,K \cite{Fontes}.
Therefore, the observation of a `magnetic` peak persisting up to 10\,K may be connected
with admixture of 1:1:1:1 phase.

Thus, the origin of the `magnetic` peak in the PC spectra has to be still under
debate and we will consider further scenarios. It could be connected with a transition
to the thermal regime by a voltage rise, which produces peak-like features
in the PC spectra at e$V_{\mbox{\tiny M}}\simeq 3.63$k$_{\mbox{\tiny B}}
T_{\textrm{\tiny M}}$ (here $T_{\textrm{\tiny M}}$ is a magnetic transition
temperature) as it was shown for ferromagnetic metals \cite{Verkin}. However,
this is not likely the case, because an increase of temperature in
the PC would drastically smear the phonon maxima in the spectra as well. On the other
hand, it was shown in  Ref.\ \onlinecite{Kulik1} that the reabsorbtion of
nonequilibrium phonons generated by electrons in a PC leads to
a lattice heating up to a temperature of e$V$/4 (and the appearance of a
background in the PC spectra), however, here the electron gas of the conduction
electrons should remain cold preventing a  smearing of the PC spectra whereas the
local Ho 4$f$ moments might be affected due to a strong enough
magnetoelastic coupling \cite{Kreyssig99,Fil}. We suppose that
such a "heating" by nonequilibrium phonons suppresses the magnetic order. The destruction of
the antiferromagnetic order in HoNi$_{2}$B$_{2}$C leads to a sharp increase of the
resistivity at $T_{\textrm{\tiny N}}$ (see, e.\,g., Fig.\,3 in
Ref.\ \onlinecite{Rathna}), which manifests itself in the PC spectra as a
peak-like feature. Indeed, the position of the peak around
2.5 meV (see Fig.\,\ref{hof4}) corresponds to the temperature
e$V{\mbox{\tiny M}}/4\simeq $ 7\,K $\sim T_{\mbox{\tiny N}}$.

\subsection{PCS of the SC energy gap in the commensurate antiferromagnetic
phase}

\begin{figure}
\begin{center}
\includegraphics[width=8cm,angle=0]{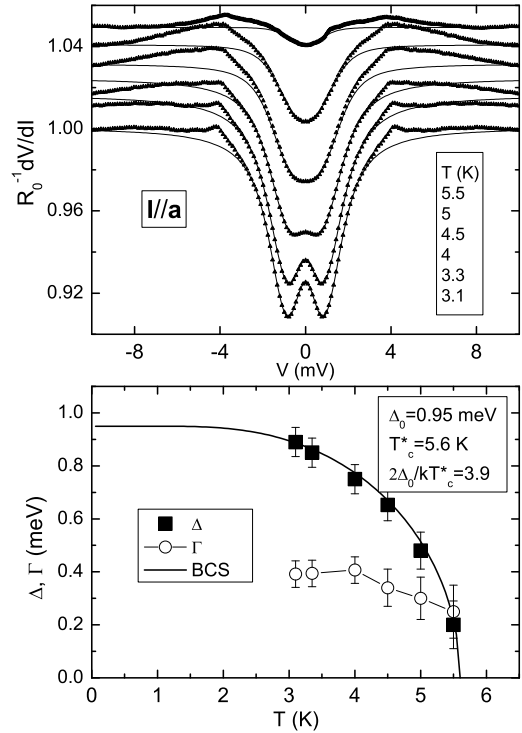}
\end{center}
\caption{Upper panel: $dV/dI$ curves (symbols) of
HoNi$_{2}$B$_{2}$C--Cu contact ($R=2.7\,\Omega$) established along
a-axis with varying temperature. PC spectra for this contact are
shown in Fig.\,\ref{hof4}(a). Solid lines are BTK fitting curves.
Bottom panel: temperature dependence of the SC energy gap $\Delta$
and `smearing` parameter $\Gamma$ at $B$=0\,T obtained by a
BTK-like fitting of the curves shown in the upper panel at
constant `tunneling` parameter $Z\approx$0.5. Solid smooth line is
the BCS curve.}
 \label{hodt}
\end{figure}

For the N-c-S metallic junctions (here N denotes a normal
metal, c is the constriction and S is a superconductor)
under consideration also the SC
energy gap $\Delta$ can be investigated (see, e.g., \cite{Naid},
Chapter 3.7). The SC gap manifests itself in the $dV/dI$
characteristic of a N-c-S contact as pronounced minima around
$V\simeq\pm\Delta$ at a temperature well below T$_{c}$. Such
$dV/dI$ curves are presented in Fig.\,\ref{hodt}. Using the BTK
equations \cite{BTK82} to fit $dV/dI$, the SC gap $\Delta$ and its
temperature dependence are established (see Fig.\,\ref{hodt}). It
is seen that $\Delta(T)$ has a BCS-type dependence, however the
temperature where the gap vanishes T$_{c}^{*}\simeq$5.6\,K is well
below T$_{c}\simeq$8.5\,K of the bulk crystal, i.e., the gap
disappears close to T$_{N}$. Similar observation of
a gap which vanishes at few Kelvin below T$_{c}$ were presented in
\cite{Ryba1,Ryba2} both for poly- and single crystals of
HoNi$_{2}$B$_{2}$C. In Ref. \cite{Ryba2} the authors supposed a
gapless state between T$_{c}^{*}$ and T$_{c}$ based on the
observation of an abrupt suppression  of SC  features in $dV/dI$
when crossing T$_{c}^{*}$ and the failure of a BTK fit above
T$_{c}^{*}$ \cite{Bobrov}. Note, that in the mentioned temperature
range HoNi$_{2}$B$_{2}$C exhibits \cite{Muller} a specific
magnetic spiral structure. According to \cite{Amici,doh} the
helical magnetic order weakens the superconductivity and even
nonmagnetic impurities do suppress superconductivity in systems
with coexisting superconductivity and antiferromagnetism
\cite{Morozov}.

We would like also to accentuate that the same gap value with a
similar temperature and magnetic field behavior has been found for
a cleaved surface of the same HoNi$_{2}$B$_{2}$C single crystal
with an Ag counter electrode, that is for some arbitrary
direction. This observation together with analogous data for
$\Delta$ from \cite{Ryba1,Ryba2} both for poly- and single
HoNi$_{2}$B$_{2}$C crystals demonstrates at least a small gap
anisotropy and a reasonable physical meaning
of T$_{c}^{*}$. Therefore we adopt the ''Fermi surface separation'' scenario for the
coexistence of magnetism and superconductivity in magnetic
borocarbides, i.e.\ their coexistence on different Fermi surface
sheets (FSS), proposed in Refs.\,\cite{drechsler01,drechsler03,Shorikov}.
In other words, we suggest that the superconductivity in the commensurate
antiferromagnetic phase survives only at a special (nearly isotropic) FSS
with no admixture of Ho 5 $d$ states. The latter mediate the magnetic
exchange interaction between the localized Ho 4$f$ moments and a part
of the delocalized conduction electrons for which a standard
$s$-wave superconductivity is easily to be destroyed or strongly suppressed.
Such a magnetic link between the Ho 4$f$ moments provided by conduction
electrons is necessary for the RKKY interactions to explain
the observed magnetic orderings at relative high temperatures.
In order to proceed with a semi-quantitative analysis of our PC data,
we may now adopt the well understood standard isotropic single band (ISB)
model to analyse the residual superconductivity. From the measured
gap value $\Delta_0$=0.95 meV and the estimated
$T_{c}^{*}\approx5.6\textrm{ K}$ (or the slightly different value
of 5.8 K as discussed below) remarkable strong coupling
corrections can be estimated. Starting for instance from a
relation first derived by Geilikman and Kresin
\cite{geilikman,carbotte}
\begin{equation}
\frac{2\Delta(0)}{{\rm k}_{\mbox{\tiny B}}T_{c}}=3.53\left[1+12.5\left(\frac{{\rm k}_{\mbox{\tiny B}}T_{c}}
{\hbar\omega_{l}}\right)^{2}\ln\left(\frac{\hbar\omega_{l}}{2{\rm k}_{\mbox{\tiny B}}T_{c}}\right)\right],
\label{delta}
\end{equation}
one estimates a relatively low value for the logarithmically
weighted phonon frequency $\hbar\omega_{l}\approx$ 10 meV. Using
the well-known Allen-Dynes expression for the critical temperature
we may estimate the effective electron-boson coupling constant
$\lambda_{s}$ as 0.934 adopting a typical value for the Coulomb
pseudopotential $\mu^{*}$=0.13. Then, from the estimated upper
critical field $H_{c2}(0)\approx$ 0.8 T one estimates in the clean
limit the effective Fermi velocity $v_{\mbox{\tiny F}}$ according to
\cite{Shulga}
 \begin{equation}
v_{\mbox{\tiny F}}\left[10^{5}\mbox{m/s}\right]=0.154\frac{T_{c}\left[\mbox{K}\right]
\left(1+\lambda\right)^{1.1}}{
\sqrt{H_{c2}(0)\left[\mbox{T}\right]}}, \label{vfermi}
 \end{equation}
as a lower bound $v_{\mbox{\tiny F}}\approx 2.06\times10^{5}$ m/s. A similar
value $v_{\mbox{\tiny F}}\approx 1.91\times10^{5}$ m/s follows, if
$H_{c2}(0)\simeq$0.93\,T is estimated from the slope of
$dH_{c2}/dT$ near $T_c$ using the well-known WHH-relation between
the former two quantities. These estimates may be helpful to
elucidate which of the FSS does bear the residual
superconductivity. Compared with the nonmagnetic borocarbides
($\lambda\sim1.2$ to 1.4) the obtained coupling constant is
surprisingly less suppressed and the low value
$\hbar\omega_{l}\approx$ 10 meV indicates that most of the phonon
modes above 10\,meV shown in Fig.\,\ref{hof2} should be regarded
as relevant for those nonactive FSS's where the superconductivity
has been extruded by the rare earth induced magnetism. Since in
the scenario under consideration the pairing comes mainly from the
interaction with low-frequency bosonic modes, the observed peak
near 10\,meV and its decomposition into contributions from
different bosons deserves special interest. In particular,
assuming that the remaining superconducting FSS couples dominantly
to the modes which contribute to the $10\textrm{ meV}$ peak, a
non-negligible contribution $\lambda_{s,\mbox{\tiny CEF}}$ from the attractive
interaction with crystal-field excitations (i.e.\ from the
inelastic asymmetric Coulomb scattering as proposed by Fulde
\cite{fulde}) would be rather challenging. Regarding the exact
frequency values of the relevant bosonic modes, the large
sensitivity of Eq.\,(\ref{delta}) to the actual $T_{c}$ value
should be stressed.
In this context the rather similar values of
$T_{c}\approx6.1\textrm{ K}$, $H_{c2}(0)\approx$ 0.75 to 0.9 T
\cite{winzer,yanson00} and 2$\Delta(0)/k_BT_{c}\approx$ 3.81 for
DyNi$_{2}$B$_{2}$C are noteworthy (see Ref.\ \onlinecite{yanson00}
Fig.6). Here $\hbar\omega_{l}\approx9.9\textrm{meV}$ and
$\lambda_{s}$= 0.939 (for $\mu^{*}=0.13$)
 can be derived from Eq.\,(\ref{delta}) and the
Allen-Dynes $T_c$-formula.

Such a similarity between HoNi$_{2}$B$_{2}$C and
DyNi$_{2}$B$_{2}$C seems to be natural because the same
commensurate antiferromagnetic phase is probed for both closely
related systems which exhibit within the LDA (local density
approximation) nearly the same electronic structure
\cite{divis}. Also similar phonon spectra are expected. In this
context, in comparing the two SC phases which coexist with the same
commensurate antiferromagnetic phase in HoNi$_{2}$B$_{2}$C and
DyNi$_{2}$B$_{2}$C, the violation of the frequently used De
Gennes-scaling for magnetic borocarbides \cite{Muller} is
striking. This scaling employs the scaling factor DG defined as
\begin{equation}
DG=\left(g_{L}-1\right)^{2}J(J+1),
\end{equation}
where $g_{L}$ is the Land\'{e}-factor and $J$ denotes the magnetic
moment at the rare earth site. According to that standard
scenario an $enhanced$ SC state would be expected for the low-temperature
SC phase of HoNi$_{2}$B$_{2}$C. However, by comparing the SC
transition temperatures and gap values of the magnetically
adjacent borocarbides contradictory results are derived.
The PC-data for ErNi$_{2}$B$_{2}$C \cite{yanson00b} yield
$\Delta(0)_{Er}\approx$ 1.7 meV (ignoring a weak anisotropy of
$H_{c2}$), the gap of DyNi$_{2}$B$_{2}$C amounts to
$\Delta(0)_{Dy}\approx$ 1.0\,meV and
$\Delta(0)_{Er}>\Delta(0)_{Ho}>\Delta(0)_{Dy}$ with
$T_{c,Er}=10\textrm{ K}>T_{c,Ho}=8.5\textrm{ K}>T_{c,Dy}=6\textrm{
K}$ is expected. This is in contrast with the measured value of
0.95\,meV for HoNi$_{2}$B$_{2}$C. Anyhow, the measured relation
$\Delta(0)_{Ho}\leq\Delta(0)_{Dy}$ suggests that the residual
superconductivity below 6\,K is not  affected by the usual magnetic pair
breaking. Our FSS-related coexistence picture allows also a new
interpretation of the specific heat data reported by Michor \textit{et al.}
\cite{michor}: using the estimates for DyNi$_{2}$B$_{2}$C given
above, strong coupling corrections for the relative specific heat
jump \cite{carbotte}
 \begin{equation}
\frac{\Delta
C}{\gamma_{s}{\rm k}_{\mbox{\tiny B}}T_{c}}=1.43\left[1+53\left(\frac{{\rm k}_{\mbox{\tiny B}}T_{c}}
{\hbar\omega_{l}}\right)^{2}\ln\left(\frac{\hbar\omega_{l}}{3{\rm k}_{\mbox{\tiny B}}T_{c}}\right)\right],
 \end{equation}
can be used to estimate the partial density of states
$\gamma_{s} \propto N_{s}(0)(1+\lambda_{s})$ of the electronic subgroup
involved in the SC transition. The observed reduced SC
specific heat jump of $\Delta C\approx$ 70 mJ/molK for
DyNi$_{2}$B$_{2}$C, that corresponds to a partial density of
states of about $N_{s}(0)=$ 4.16 mJ/molK$^{2}$, only, to be compared
with about $N_{tot}(0)=$ 10 mJ/molK$^{2}$ for the total density of
states as derived from LDA band structure calculations for
nonmagnetic and magnetic borocarbides \cite{divis}.

It is interesting to compare $\lambda_{\mbox{\tiny PC}}$ with the
superconducting coupling constant $\lambda_s$ derived above as
well as with superconducting and high-temperature resistivity data
which yield $\lambda_{tr}$ for related systems in order to
understand to which extent $\lambda_{\mbox{\tiny PC}}$ probes the el-ph
coupling for all bands.
Recent PC measurements for DyNi$_{2}$B$_{2}$C \cite{yanson00a} and
YNi$_{2}$B$_{2}$C \cite{Bashlakov06} in the best junctions yield a
significant larger (smaller) experimental coupling constant
$\lambda_{\mbox{\tiny PC}}=0.4$ (0.1), respectively, compared to the present
case with $\lambda_{\mbox{\tiny PC}}=0.25$ for HoNi$_{2}$B$_{2}$C. However,
very often the coupling constants derived from point-contact
spectra should be considered as lower bounds for the coupling
constants relevant for superconductivity, say $\lambda_{\mbox{\tiny PC}}\approx $
(0.5 to 0.7)$\lambda_s$, which can be derived from different Eliashberg
functions $\alpha^2(\omega)F(\omega)$ extracted by the inversion of usual superconductor
tunneling data or from an analysis of the Sommerfeld constant of the linear specific
heat in the normal state. Since a value of
$\lambda_s\sim1.2$ for the case of the non-magnetic borocarbides
and $\lambda_{tr}\approx$ 0.8-0.9 for magnetic borocarbides
\cite{Rathna} seems to be realistic, we adopt for ideal ballistic
contacts $\lambda_{\mbox{\tiny PC}}\sim\lambda_{tr}\approx$ 0.7 to 0.8 for all
Ni-1221 borocarbides under consideration. In fact, the coupling
constants, measured by point-contact spectroscopy in the normal
state can be regarded as effective single-band properties, i.e.\
for the multiband system under consideration generalizing Eq.\
(1), each band (FSS) produces a weighting factor given by the
ratio of the corresponding conductances and the Fermi velocities
$v_i$ which in a free electron-like picture reads
\begin{equation}
\lambda_{\mbox{\tiny PC}}=\frac{v_{eff}}{\sigma_{tot}}\sum_i
\left[\frac{\sigma_i}{v_i}\lambda_{{\mbox{\tiny PC}},i}\right], \label{lambda}
\end{equation}
where the band index `$i=$1,.. has been introduced. In the present
case following our coexistence scenario we  divide the
borocarbides under consideration into a magnetic ($i$=1) and a
superconducting subsystem ($i=$2) with $v_{eff}\approx 1.8v_2$
used in calibrating the point-contact spectra and adopt $v_1\sim 3
v_2$, as suggested by the LDA-FPLO calculations, one estimates
$\sigma_{i}/\sigma_{tot}$ =0.9 and 0.1, respectively. Then Eq.\
(8) can be rewritten as $\lambda_{\mbox{\tiny PC}}=0.18(3\lambda_1+\lambda_2)$.
For $\lambda_1 \sim \lambda_2$ a moderate contribution of about
25\% for the superconducting subsystem to the total $\lambda_{\mbox{\tiny PC}}$
follows. The latter is dominated by the quenched magnetic
subsystem for HoNi$_{2}$B$_{2}$C and other magnetic borocarbides.
Anyhow, the 10 meV peak can be attributed completely to the
residual superconducting subsystem.  A similar result was obtained
for a specific heat analysis of HoNi$_{2}$B$_{2}$C \cite{waelte},
where the peak near 10\,meV was estimated to contribute about 25\%
to the total coupling constant of 1.2. Related problems and
more detailed further comparisons with DyNi$_{2}$B$_{2}$C, ErNi$_{2}$B$_{2}$C,
and other magnetic borocarbides will be considered elsewhere.

For completeness and for comparison with Eq. (\ref{lambda}) we
show the corresponding multiband expressions for the
renormalization of the electronic specific heat (the Sommerfeld
constant $\gamma$) and the resistivity
\begin{equation}
\lambda_{\gamma}=\sum_i\frac{N_i(0)}{N_{tot}(0)}\lambda_{\gamma i}~~,
\end{equation}
where $N_i(0)$ denotes the partial electronic density of states at the Fermi
level. Finally, ignoring interband scattering for the high-temperature resistivity
where $\rho(T) \propto \lambda_{tr}T$ one obtains from a straightforward
generalization of the well-known Ziman-formula
\begin{equation}
\lambda_{tr}=
\left( \sum_i \frac{\omega^2_{pl,i}}{\Omega^2_{pl,tot}}\frac{1}{\lambda_{tr,i}}\right)^{-1},
\end{equation}
where $\omega_{pl}$ denotes the plasma frequency. Thus, in all
discussed cases the effective one-band coupling constants are differently
decomposed into the partial coupling constants depending on which
quantity is probed experimentally.

\begin{figure}
\begin{center}
\includegraphics[width=8cm,angle=0]{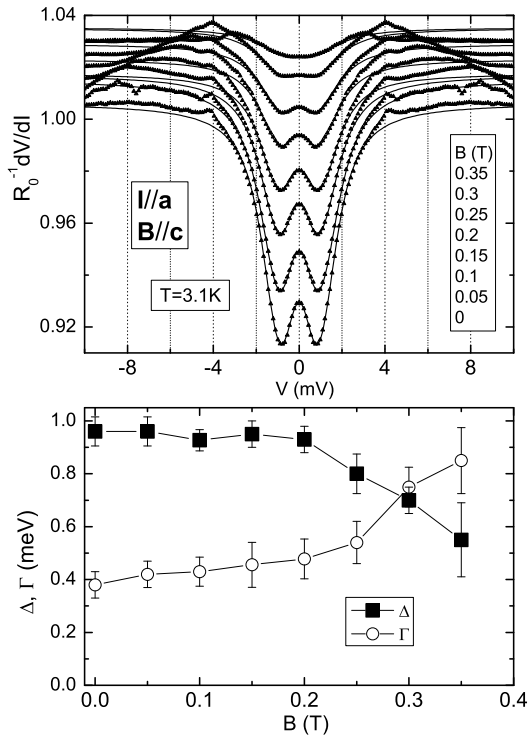}
\end{center}
\caption{Upper panel: The $dV/dI$ curves (symbols) for the same
HoNi$_{2}$B$_{2}$C--Cu contact as employed in Fig.\, \ref{hodt}
but exposed to an external magnetic field. The solid lines are
BTK fitting curves. Bottom panel: the magnetic field dependencies
of the SC energy gap and $\Gamma$ at constant $Z\approx$0.5 obtained
by BTK-like fits  of the curves shown in the upper panel.}
 \label{hodh}
\end{figure}

\begin{figure}
\begin{center}
\includegraphics[width=8.5cm,angle=0]{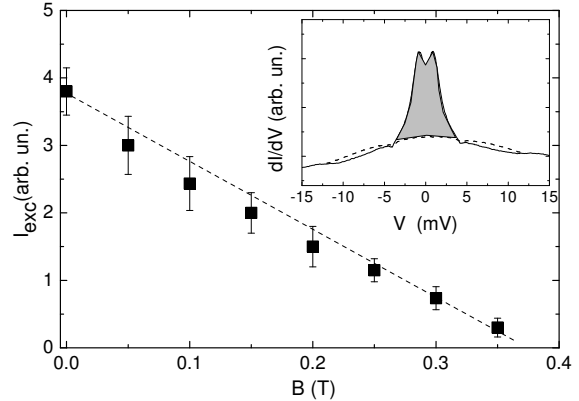}
\end{center}
 \vspace{0cm}
\caption{Excess current $I_{\textrm{exc}}$ behavior versus magnetic
field (symbols) for PC data used also in Fig.\,\ref{hodh}. Dashed straight line
connects two edge symbols. The inset explains the determination of the excess
current, which is proportional to the integral intensity of $dI/dV$
maximum (the shaded area) after subtracting of the normal state $dI/dV$
(dashed curve) measured at 0.5\,T. }
 \label{iex}
\end{figure}

Lets turn to the magnetic field measurements shown in
Fig.\ref{hodh}. In contrast to the PC data reported
in Ref.\ \onlinecite{Ryba2}, where $\Delta$ initially increases ($\sim10\%$) in
magnetic fields before it drops off to zero, in our case $\Delta(B)$
exhibits a conventional behavior. decreasing with overall negative
curvature (see Fig.\,\ref{hodh}). Similarly, the excess current
$I_{\textrm{exc}}$, which is proportional to the gap maximum area
in $dI/dV$ (see Fig.\,\ref{iex} and the figure caption therein), exhibits a
temperature dependence (not shown) like $\Delta(T)$ from
Fig.\,\ref{hodt}, that is $I_{\textrm{exc}}(T)$ has also a
negative curvature. (We remind the reader, that the excess current
of an S-c-N contact is governed by $\Delta$ \cite{Artemenko} or
the SC order parameter \cite{Belob}.) Contrary,
$I_{\textrm{exc}}(B)$ decreases nearly linearly with the magnetic
field (Fig.\,\ref{iex}). This is in contrast with the observed
pronounced positive curvature of $I_{\textrm{exc}}(B)$ in
MgB$_{2}$ \cite{NaidyukF} and YNi$_{2}$B$_{2}$C \cite{Bashlakov},
where at least two bands are superconducting. Thereby,
the observed behavior of $I_{\textrm{exc}}(B)$ is in line with our scenario
that superconductivity survives on separate (one or similar) FSS.
In addition to the "vortex" model proposed in \cite{NaidyukF},
the $I_{\textrm{\small exc}}(B)$ dependence
might be also related  to a complex vortex structure for a system
with normal and superconducting FSS's as we suppose in our case.
To the best of our knowledge this vortex problem has not been
studied so far theoretically.

\section{Conclusion}

We have carried out investigations of the EBI spectral function
and the SC energy gap in HoNi$_{2}$B$_{2}$C by PC spectroscopy. We
succeeded to measure PC EBI spectra of HoNi$_{2}$B$_{2}$C with
distinct phonon features above 15\,mV along with additional low
energy maxima of non-phonon nature below 15\,meV. For the first
time the high energy maxima at about 50\,meV and 100\,meV are
clearly observed. Hence, the high energy phonons can not be
disregarded by analyzing thermodynamic, electron transport and SC
phenomena in the nickel borocarbides. The phonon maxima
in the PC spectra correspond well to those in the phonon DOS spectra of borocarbides
measured by inelastic neutron scattering. On the other hand the maximum at
about 10\,meV contains likely also an essential contribution from CEF
excitations, while the low-lying peak around 3\,meV is suppressed
by temperature and magnetic field rise. Moreover, the peak intensity depends
on the angle between the magnetic field and the
$c$-axis. All these observations point to a `magnetic` origin of this peak,
i.\,e., the peak reflects possibly a transition from the low
temperature magnetically ordered state to dis- (or low)-
ordered states driven by nonequilibrium phonons generated by
accelerated electrons.

The SC gap exhibits in general a `conventional` single-band like standard
BCS behavior below T$_{c}^{*}\simeq5.6$\,K with
$2\Delta/{\rm k}_{\mbox{\tiny B}}$T$_{c}^{*}\simeq3.9$ pointing to an intermediately
strongly coupled SC state.  The large similarity of this
low-temperature phase with the behavior of DyNi$_{2}$B$_{2}$C
where superconductivity and a simple commensurate
antiferromagnetic state coexist, clearly points to a nonstandard
(in the sense of the frequently used De Gennes scaling)
suppression of the former compared with nonmagnetic
borocarbides. We addressed the specific role of selected Fermi
surface sheets and CEF-excitations in this particular case.
Between T$_{c}^{*}$ and the upper critical temperature
T$_{c}\simeq$8.5\,K the SC signal in $dV/dI$ is drastically
suppressed, which explains the difficulties to derive the SC order
parameter (or a SC gap) by fitting the $dV/dI$ data. Hence the SC
state between T$_{c}^{*}$ and T$_{c}$, in the region of peculiar
magnetic order, is seemingly `unconventional`. Due to its weakness
there is a large sensitivity to both external effects (stress or
pressure by PC creation, high current density in PC) and to the
intrinsic properties (small deviation from stoichiometry, a
specific SC state at the surface caused by, e.g.,\ peculiar
magnetic ordering). Further investigations are desirable to settle
the nature of the SC state in this special temperature range to
obtain finally a similar level of understanding as for
the low-temperature SC phase coexisting with the commensurate
antiferromagnetic state we have achieved here.
In a more general context one may only agree with the conclusion drawn in a
very recent review paper \cite{Budko06} as to
''the $R$Ni$_2$B$_2$C series will continue to provide insight to
superconductivity and its interaction and response to local moment magnetism
for years, and possibly decades to come.''

\section*{Acknowledgements}

The support by the Deutsche Forschungsgemeinschaft (SFB 463) and the
State Foundation of Fundamental Research of Ukraine are acknowledged.
One of the authors (O.E.K.) was supported by the DAAD through a NATO-Grant.
We are indebted to L.~V.~Tyutrina for the help in fitting the $dV/dI(V)$
data and W.\ L\"{o}ser for a critical reading of the manuscript.
Discussions with K.-H. M\"{u}ller, M.\ Schneider and H.\ Rosner are acknowledged.


\begin{thebibliography}{10}
\bibitem{lambda} In discussing the value of $\lambda_{\mbox{\tiny  PC}}$ derived
 from a PC spectrum, we should always to bear in mind that it is a
transport coupling constant which in general differs from
the thermodynamic counter parts $\lambda_s$, although
for some metals (see e.g.\ Ref.\ \onlinecite{Naid})
$\lambda_{\mbox{\tiny  PC}} \approx  \lambda_s$
holds accidentally.
\end{thebibliography}
\end{document}